\documentclass[twocolumn,pre]{revtex4-1}
\usepackage{graphicx}
\usepackage{amsmath}
\usepackage{amstext}
\usepackage{amsfonts}
\usepackage{diagbox}
\usepackage{setspace}

\newcommand{\bolde}{\mathbf{e}}

\newcommand{\boldb}{\mathbf{b}}
\newcommand{\boldB}{\mathbf{B}}

\newcommand{\om}{\omega}

\newcommand{\bu}{\mathbf{u}}
\newcommand{\boldI}{\mathbf{I}}
\newcommand{\boldL}{\mathbf{L}}
\newcommand{\be}{\begin{equation}} 
\newcommand{\ee}{\end{equation}} 
\newcommand{\bea}{\begin{eqnarray}} 
\newcommand{\eea}{\end{eqnarray}} 
\newcommand{\bc}{\begin{center}} 
\newcommand{\ec}{\end{center}}

\begin{document}

\title{Explicit construction of the eigenvectors and eigenvalues of the graph Laplacian on the Cayley tree}

\author{ Ay\c{s}e Erzan$^1$ and Asl{\i} Tuncer$^2$}
\affiliation{$^1$ Department of Physics, Istanbul Technical University, Maslak 34469, Istanbul, Turkey}
\affiliation{$^2$ Department of Physics, Ko\c{c} University,  Sar{\i}yer 34450, Istanbul, Turkey}

\begin{abstract}
A generalized Fourier analysis on arbitrary graphs calls for a detailed knowledge of the eigenvectors of the graph Laplacian.  Using the symmetries of the Cayley tree, we recursively construct the family of eigenvectors with exponentially growing eigenspaces, associated with eigenvalues in the lower part of the spectrum. The spectral gap decays exponentially with the tree size, for large trees.  The eigenvalues and eigenvectors obey recursion relations which arise from the nested geometry of the tree. Such analytical solutions for the eigenvectors of non-periodic networks are needed to  provide a firm basis for the spectral renormalization group which we have proposed earlier [A. Tuncer and A. Erzan, Phys. Rev. E {\bf 92}, 022106 (2015)].


\end{abstract}
\date{\today}
\maketitle

\section {Introduction}
The eigenvectors and the eigenvalues of the graph Laplacian, on networks lacking translational invariance, have gained new relevance with the introduction of spectral methods~\cite{Cassi,Jost2009,Dorogovtsev2003,Chung2,Newman2003,Bianconi2010} into  study of the topological and dynamical properties of 
arbitrary networks~\cite{Dorogovtsev2002,Barabasi2002,Dorogovtsev2008}. 

In an earlier paper~\cite{Tuncer2015} we have formulated a renormalization group for a polynomial field theory living  on  networks which are not translationally invariant and not necessarily embedded in metric spaces.  The formulation is based on a generalized Fourier transform using the eigenvectors of the graph Laplacian.  The precise form of the eigenvectors do not come into the computations for the Gaussian model; however they are crucially important when higher order interactions, such as $\psi^4$, are included.

Here we will examine the eigenvectors and eigenvalues of the graph Laplacian~\cite{Chung} for the Cayley tree, a network which is very rich in symmetries and lends itself to many interesting mathematical and physical applications.  The eigenvectors for $0\le \om \le 1$ have an interesting recursive structure which follows from the symmetries of the Cayley tree, and also incorporate lower dimensional periodic properties. These features are also evident in the structure of the graph Laplacian, which we express in a compact form, making explicit use of these symmetries. 

In the study of phase transitions and critical fluctuations on arbitrary networks, it is the lower part of the spectrum (corresponding to the relatively longer wavelengths on a periodic lattice) which is of greater significance.  
For the Cayley tree, we already know from our numerical computations~\cite{Tuncer2015}  that,  at $\omega=1$, we encounter a feature  analogous to the Van Hove singularity.  Evidence of Van Hove singularities have been encountered in other lattices without long range  periodic translational 
order.~\cite{Choy} In this paper we will mainly confine our analysis of eigenvalues and eigenvectors to the interval $0\le \om \le 1$, where the spectral density exhibits an exponentially increasing trend, with a kink at $\om=1$.  The  ``surface" to volume ratio of a regular Cayley tree with branching number $b$ for large $r$ is asymptotically equal to $(b-1)/b$, and therefore finite size effects are always strongly present. 

Ochab and Burma~\cite{Ochab} have presented derivations of the eigenvalues and eigenvectors of the adjacency matrix of the Cayley tree (also see Zimmermann and Obermaier~\cite{Zimmer}).  Rojo {\it et al.} have given a set of implicit matrix equations~\cite{Rojo,Rabbiano} for the eigenvectors of the graph Laplacian. However, their approach does not provide sufficient insight into the structure of the eigenvectors. For comparison, also see Malozemov and Teplyaev~\cite{Malozemov}, who derive an iterative formalism for the spectrum of deterministic substitutional graphs.

In Sec. II  we establish the notation, we recall the symmetries of the Cayley tree, provide an iterative construction for the graph Laplacian and compute the eigenvalues. In Sec. III  we present an explicit construction of the first $r$ eigenspaces and the relevant eigenvectors of a Cayley tree with $r$ generations of nodes, making use of the symmetries of the graph.  
Sec. IV provides a brief discussion. 

\section{The symmetries of the Cayley tree, the graph Laplacian and eigenvalues}

A graph is a set of nodes, which may be pairwise connected by edges.  In this paper we will only be concerned with undirected edges, and only one edge, if any, is permitted between each pair of nodes.  A tree is a graph whose edges do not form any loops.  Here we will only deal with one connected tree, with $b$ edges emanating from an initial node, and uniform branching number $b$.   The initial node we call the ``root", and define as the 1st generation. The nodes that are connected to the root by a path consisting of $s-1$ edges constitute the $s$th generation. A Cayley tree with $r$ generations will be called an $r$-tree. 

The degree $d$ of a node is the number of nodes to which it is connected by an edge. We will designate the set of nodes that are mutually attached to a node on the previous generation as a ``shell" and those nodes which reside on the outermost generation of the tree we will call ``leaves."   
\subsection{The primitive tree as the generator of the Cayley tree}

We  define the ``primitive tree'' as consisting of $b$ nodes attached to a root node,  thus having $N=b+1$ nodes in all. 
This is a ``2-tree,'' having a root (which belongs to the first generation) and a second generation which is just a shell consisting of leaves. (Note that we have changed the convention for numbering the generations from Ref.~\cite{Tuncer2015}). 
 
Define the  ``multiplication" $Y\ast X$ of two trees, say $Y$ and $X$, as attaching $Y$ by its root to all the leaves of $X$. An $r$-tree can be obtained by repeated ``multiplications" of the primitive tree with itself. Thus a uniform  $r$-tree may be written as a  $Y(b)^{r-1}$, the $r-1$st ``power" of a primitive tree with branching number $b$. (See Appendix 1.)
\subsection{Symmetries of the Cayley tree}
The generator of the symmetry group of the Cayley tree with branching number $b$  is the permutation group $G_b$ of $b$ objects and the symmetry group of the $r$-tree is simply $G^{N_r}_b$, where $N_r=(b^r-1)/(b-1)$ is the number of nodes on the tree.  

Define a subtree as a part of the whole tree, emanating from an arbitrary node on the $(1< s)$th generation.  The subtree $Y(b)^{r-s+1}$ also has the permutation symmetry about each of its nodes. This leads to a scaling relation between  the symmetry group of $Y(b)^{r}$ and that of  $Y(b)^{r-s+1}$, given by $G_b^{N_r-N_s}$.
\subsection{The graph Laplacian}
The graph Laplacian $\mathbf{L}$ is defined~\cite{Chung} as a matrix operator, $L_{ij}=\delta_{ij}d_i-A_{ij}$ where $d_i$ is the degree of the $i$th node, $i,j=1,\ldots N$, $\delta_{ij}$ is the Kronecker delta and $A_{ij}=1$ if the nodes $(i,j)$ are connected by an edge of the graph and zero otherwise. From this it is clear that the constant vector (with all elements equal to each other) is an eigenvector with null eigenvalue and, since the tree is connected, this eigenvalue is non-degenerate. All the other eigenvalues are larger than zero. 

It is customary to label the eigenvalues of the graph Laplacian in increasing order, $0=\om_1 \le \om_2, \ldots \le \om_N$ for a graph with $N$ nodes \cite{Chung}.
We will write $\om^{(r)}_{(n)}$,  to refer to the $n$th distinct eigenvalue for an $r$-tree. A particular eigenvector within the $n$th eigenspace will be denoted by $\mathbf{u}^{(r)}_{\ell, (n)}(\rho)$.  The index $\ell$ is the analogue of a wave number and $\rho$ indicates the node where the subtree, spanning the nonzero elements of the eigenvector, is rooted. When it is obvious from the context, one or more of these indices may be omitted but we will keep the parenthesis around the last subscript so that there is no ambiguity.
\subsubsection{The Laplacian, its eigenvectors and eigenvalues for the primitive tree}
Define $\mathbf{b}$ to be a column vector with $b$ elements, all of them unity, $\mathbf{b}^\dagger = (11...1)$,  and $\mathbf{I}_b$ is the identity matrix of $b$ dimensions. The $(b+1) \times (b+1)$ Laplacian matrix for the primitive tree ($r=2$) with branching number $b$ can then be written as,
\be
\boldL^{(2)} =
\begin{pmatrix}
b& -{\mathbf b}^\dagger\\
-{\mathbf b} & \mathbf{I}_b\\
\end{pmatrix}
\label{primL}
\ee 

The constant vector, with all $b+1$ elements equal, is an eigenvector which we will denote by  $\mathbf{u}^{(2)}_{(1)}$, and trivially gives rise to the first eigenvalue $\omega^{(2)}_{(1)}=0$ with  degeneracy $\tau^{(2)}_{(1)} =1$. (Note that $\om^{(r)}_{(1)}=0$, with the eigenvector being the constant vector, is true for all $r$, since the rows and the columns of the graph Laplacian have zero sums. This also ensures that the sum over the elements of each eigenvector is zero.) 


Define the subvector, 
\be
 \bolde^{(\ell)} \equiv \begin{pmatrix}  e^{(\ell)}_1\\e^{(\ell)}_2\\ \ldots\\ e^{(\ell)}_b\\ \end{pmatrix} 
\ee
where $e^{(\ell)}_j= e^{i \theta_j\ell}$ where $\theta_j=2j\pi/b$ and $\ell=1, \ldots, b-1$.  Note that $2\pi/ (b/\ell)$ is the analogue of a wave number.  On the other hand $p$ successive $b$-cyclic permutations of the elements of the vector $\bolde^{(\ell)}$, for a given $\ell$,  will only multiply this vector uniformly by a phase $e^{i2\pi p\ell /b}$ and therefore do not yield an independent vector.  

The vectors 
\be
 \bu^{(2)}_{\ell, (2)} = \begin{pmatrix}0\\\bolde^{(\ell)} \end{pmatrix} \;\;.
\ee
make up the eigenspace yielding the smallest nonzero eigenvalue $ \om^{(2)}_{\ell, (2)}=1$  for the primitive tree, with degeneracy $\tau^{(2)}_{(2)}=b-1$.

The eigenvector $(-b, 1, \ldots, 1)$ (up to normalization) gives the last eigenvalue $\om^{(2)}_{(3)}=b+1$. 
\subsubsection{Graph Laplacian for $r$-trees}

We may now write down the graph Laplacian for an  $r$-tree having uniform branching number $b$, using the so called direct (Kronecker) product. Ochab {\it et al.}~\cite{Ochab} also represent the adjacency matrix in block form, but without the advantage of our compact notation which makes the graph symmetries evident.

The Kronecker product  $\mathbf{A} \otimes \mathbf{B}$  of two matrices is formed by multiplying the matrix $\mathbf{B}$ from the left by each element $A_{kl\ldots}$ of the matrix $\mathbf{A}$.  With $\mathbf{I}_b^{(1)}\equiv \mathbf{I}_b$ being the identity matrix in $b$ dimensions, we define the $b^n$ dimensional identity matrix,
\be
{\mathbf I}_b^{(n)} \equiv \underbrace{{\mathbf I}_b\otimes {\mathbf I}_b\otimes \ldots \otimes {\mathbf I}_b}_\text{$n$ times}\;\;.
\label{Ib}
\ee
and set ${\mathbf I}_b^{(0)}=1$, a scalar, for completeness.

We have already defined  ${\mathbf b}$ in the previous subsection, and ${\mathbf b}^\dagger=(1\,1\,\ldots\,1)$.  We now define the $b^{n+1}\times b^n$ and $b^{n}\times b^{n+1}$ rectangular matrices,  ${\mathbf B}_n$ and ${\mathbf B}_n^\dagger$, $n\ge 0$ as,
\be 
{ \mathbf B}_n = {\mathbf I}_b^{(n)} \otimes {\mathbf b} \;\;\;\;{\rm and}\;\;\;\;\;
{\mathbf B}_n^\dagger={\mathbf I}_b^{(n)} \otimes {\mathbf b}^\dagger \;\;,
\label{Bnd}
\ee
where $ {\mathbf B}_0={\mathbf b}$ and ${\mathbf B}_{-1}$ is not defined. 

The ``elements'' of the graph Laplacian, ${\mathbf L}_{m,n}^{(r)}$, $m, n =1, \ldots, r$ are written in boldface, since each such element is actually a square (for $m=n$) or a rectangular (for $ m\ne n$) matrix. Each row (column) index refers to a generation, with the root having the index unity. The rectangular matrices connect the nodes in the $m$th with the $m\pm1$st generations. The degrees take the values $d_1=b$; $d_m=b+1$ for all $1<m<r$ and $d_r=1$. 

For $m=1$,
\be
{\mathbf L}_{1,n}^{(r)} = \delta_{1,n}b-\delta_{1,n-1} {\mathbf B}_0^\dagger \,\,.
\ee

For $1< m <r$, 
\be
{\mathbf L}^{(r)}_{m,n} = \delta_{m,n}\, (b+1) \,{\mathbf I}_b^{(m-1)}\, -\,  \delta_{m-1, n}\,{\mathbf B}_{n-1} \, -\, \delta_{m+1,n}\,{\mathbf B}_{n-2}^\dagger \;\;.
\label{Lfor1mr}
 \ee

For $m=r$,
\be
{\mathbf L}^{(r)}_{r,n} = \delta_{r,n}\,{\mathbf I}_b^{(r-1)}\, -\,  \delta_{r-1, n}\,{\mathbf B}_{n-1} \,\,. 
\ee
%
\subsection{The eigenvalues of the Laplacian}
Reducing the matrix $\boldL^{(r)} - \omega \boldI^{(N_r)}$ to lower triangular form, we find that the diagonal elements of this triangular matrix are actually an array of matrices $\boldI_b^{(n)}$ of dimension $b^n$, $n=0, \ldots, r-1$, with coefficients $a_r, \ldots, a_1$, satisfying the following recursion relations,
\begin{align}
a_r& = 1-\omega\\ \nonumber
a_{r-(n+1)} &= b+1-\om-b/a_{r-n}\\
a_1 & = b-\om -b/a_2  \;\;. \\ \nonumber 
\end{align}
(See Ref.~\cite{Ochab}, for a similar treatment of the adjacency matrix rather than the graph Laplacian.)
The determinant is therefore 
\be
D\equiv \det\big(\boldL^{(r)}-\om\boldI^{(N_r)}\big) = \prod_{j=1}^r(a_j)^{b^{j-1}}\;\;.
\label{D1}
\ee
It is convenient to define the polynomials $A_1=a_r, \;\; A_2=a_ra_{r-1},\;\; \ldots,  A_r=\prod_{j=r}^1 a_j$, in terms of which 
\be
D =  \bigg(\prod_{s=1}^{r-1} A_s^{b^{r-s}-b^{r-(s+1)}}\bigg) A_r \;\;.
\label{D2}
\ee
Setting the determinant equal to zero requires any one or more of the $A_s = 0$, with $1\le s \le r$. All the different eigenvalues of the Laplacian can be found in this way.  The degeneracies are given by the powers of the polynomials $A_s$ appearing in Eq.~(\ref{D2}).  

With the initial conditions $A_0=1$ and $A_1= a_r$, and $s=1, \ldots\ r-1$, the $A_s$ satisfy the following iterative equations, 
\be
A_{s} = \phi A_{s-1} \,-\,b A_{s-2}\;\;,
\label{Arecursion}
\ee
where we have defined $\phi=b+1-\om$. The recursion relation for $A_r$, however, is given by,
\be
A_r= (b-\om) A_{r-1} -b A_{r-2}\;\;.
\label{Arrecursion}
\ee

The recursion relation, Eq.~(\ref{Arecursion}), can be expressed in the form,
\be
\begin{pmatrix} A_s \\ A_{s-1} \end{pmatrix} = \cal{A} \begin{pmatrix} A_{s-1} \\ A_{s-2} \end{pmatrix} \;\;,
\label{Armatrix}
\ee
where we have defined the matrix $\cal{A}$ as,
\be 
\cal{A} \equiv \begin{pmatrix} \phi & -b  \\ 1 & 0 \end{pmatrix}\;\;.
\label{calA}
\ee

Using the Cayley-Hamilton theorem~\cite{Kahn}, we can write  ${\cal{A}}^s$ in terms of the eigenvalues of the matrix $\cal{A}$,
namely $\lambda_{1,2}= (\phi \pm \sqrt{\phi^2-4b})/{2}$, 
as,
\be
{\cal{A}}^s={\alpha_s \cal{A}}+\beta_s \boldI_2\;\;,
\ee
where 
\be
\alpha_s=(\lambda_1^s-\lambda_2^s)/(\lambda_1-\lambda_2)
\ee
and 
\be 
\beta_s=(\lambda_1^s \lambda_2- \lambda_1\lambda_2^s)/(\lambda_2-\lambda_1)\;\;. 
\ee
We get, for $1\le s\le r-1$
\be
\begin{pmatrix} A_{s}\\ A_{s-1} \end{pmatrix} = {\cal A}^{s-1} \begin{pmatrix} A_1 \\ A_0 \end{pmatrix} \;\;.
\label{Armatrix}
\ee
Setting $A_s=0$, we obtain  a nonlinear equation of the $s$th degree, to be solved for $\om$,
\be
(1- \om)= b \frac{\lambda_1^{s-1}- \lambda_2^{s-1}}{\lambda_1^{s}- \lambda_2^{s}}\;\;.
\label{omsolution2}
\ee

To obtain a similar equation for  $A_r$, we define the matrix $\cal{A}^\prime$, where we set $\phi^\prime = \phi-1=b-\om$, 
\be 
\cal{A}^\prime \equiv \begin{pmatrix} \phi^\prime & -b  \\ 1 & 0 \end{pmatrix}\;\;.
\label{calAprime}
\ee
so that 
\be
\begin{pmatrix} A_{r} \\ A_{r-1} \end{pmatrix} = {\cal A}^\prime {\cal A}^{r-2} \begin{pmatrix} A_1 \\ A_0 \end{pmatrix} \;\;.
\label{Arrmatrix}
\ee
Once again using the Cayley-Hamilton theorem finally yields, after some algebra and setting $A_r=0$, the following equation,
\be
1-\om =  b \frac{(\lambda_1^{r-1}- \lambda_2^{r-1}) - (\lambda_1^{r-2}- \lambda_2^{r-2})}{(\lambda_1^{r}- \lambda_2^{r}) - (\lambda_1^{r-1}- \lambda_2^{r-1}) }\;\;.
\label{omsolutionr}
\ee
Equations (\ref{omsolution2},  \ref{omsolutionr}) cannot be solved analytically and we will present the numerical results in the following subsection. However, we would like to make a few statements here.

The non-degenerate null eigenvalue, $\om^{(r)}_{(1)} =0$ arises in the solution of $A_r=0$, and we already see from Eq.~(\ref{D2}) that all other solutions of this equation are non-degenerate as well.
The smallest solution greater than zero, the so called spectral gap, is, in our notation, $\om^{(r)}_{(2)}$ for the given value of $r$; it is given by the smallest solution of $A_{r-1}=0$, and has a degeneracy of $b-1$. Successive eigenvalues in the ascending series, $\om^{(r)}_{(n)}$ are found as the smallest solutions of the polynomials $A_{r-n+1}$.
The degeneracies of the $2\le n \le r$th eigenspaces, namely the ascending series of eigenvalues are~\cite{Tuncer2015}, 
\be
\tau^{(r)}_{(n)}= (b-1) b^{n-2}
\label{tau}
\ee
%
%
\begin{table*}[!ht]
\label{Cayleyeigenvalues}
\caption{Eigenvalues of the Cayley tree for $b=3$, $r=1, \ldots, 9$. The vertical columns give $\om^{(r)}_{(n)}$, $1\le n \le r$.  For $r=6,\ldots,9$, we have removed from this list, the eigenvalues which are part of the large $\om>\om^\ast$  series of solutions, but nevertheless start invading the smaller $\om$ interval.  The degeneracies of the $\om^{(r)}_{(n)}$ are $\tau_{(n)} =(b-1) b^{n-2}$ for $n\ge 2$.}

\begin{center}
\begin{tabular}{|l| c c c c c c c c c|}
\hline\hline
\diagbox{$n$}{$r$} \quad  & 2 & 3 & 4 & 5 & 6 & 7 & 8 & 9 &10   \\
\hline
1&0&0&0&0&0&0&0&0& 0\\
2&1&0.20871&0.05718&0.01751&0.0056281&0.0018477&0.00061209& 0.00020353 &$6.78\times 10^{-05}$\\
3&\quad &1&0.20871&0.05718&0.01751&0.0056281& 0.0018477&0.00061209&0.00020353\\
4&\quad&\quad&1&0.20871&0.05718&0.01751&0.0056281 &0.0018477&0.00061209\\
5&\quad&\quad&\quad&1&0.20871&0.05718&0.01751&0.0056281& 0.0018477\\
6&\quad&\quad&\quad&\quad&1 &0.20871&0.05718&0.01751&0.0056281\\
7&\quad&\quad&\quad&\quad&\quad&1&0.20871&0.05718&0.01751\\
8&\quad\quad&\quad&\quad&\quad&\quad&\quad&1&0.20871&0.05718\\
9&\quad&\quad\quad&\quad&\quad&\quad&\quad&\quad&1&0.20871\\
10&\quad\quad\quad&\quad&\quad&\quad&\quad&\quad&\quad&\quad&1\\
    \hline\hline
    \end{tabular}
  \end{center}
 \end{table*}
%

We have already seen in subection II.C.1, 
that the degeneracy  $(b-1)$ comes from the ``wavenumber" $\ell \in (1,b-1)$.
In Section III, where we explicitly construct  the eigenvectors, it will become clear how the degeneracy of $b^{n-2}$ has a purely geometric meaning.  The eigenvalue $\om^{(r)}_{(r)} = 1$, with the largest degeneracy of $\tau^{(r)}_{(r)}= (b-1) b^{r-2}$, is where the  Van Hove-like singularity, or kink, arises in the spectrum.

\subsection{Numerical results}

We gain useful information from a consideration of the numerical results, which are summarised in Table I, 
where we display the numerical results for the eigenvalues in the ascending series within the interval $0\le \om \le 1$ of the spectrum and their degeneracies, for the case of $b=3$. In Appendix 2 we provide all the eigenvalues with their degeneracies, for $r\le 6$ with a brief discussion.

In Table I we see that the smallest nonzero eigenvalues for trees of increasing size are inversely ordered, with, $\om^{(r)}_{(2)} < \om^{(r-1)}_{(2)}\ldots \le  \om^{(2)}_{(2)}=1$.  Moreover, the sequence of smallest nonzero eigenvalues of trees of size 
$r-1, \ldots, 4, 3, 2$ are in fact identical with  the 3rd, 4th, $\ldots, r$th  eigenvalues of a tree of size $r$. In particular, we find that for $2\le n\le r$,
\be
\om^{(r-n+2)}_{(2)}=  \om^{(r)}_{(n)}\;\;\,
\ee
thus the series of ascending eigenvalues of smaller trees show up again in the spectrum of the larger trees, but are only shifted by one rank in the ascending order, with the insertion of a yet smaller  $\om^{(r+1)}_{(2)}$.

If we expand  the RHS in Eq.~(\ref{omsolution2}) in small  $\om^{(r)}_{(2)}$ for large $r$, we find that to lowest order,
\be
 \om^{(r)}_{(2)} \simeq (b-1) b^{-r}\;\;.
\label{asymptotic}
\ee

It is immediately clear that $\om_{(2)}^{(r)}$ depends only on $r$.
One also  sees that $\om^{(r)}_{(n)}$ are exactly coincident with $\om^{(r+m)} _{(n+m)}$, for integer $m \ge 0$. 

In Ref.\cite{Tuncer2015} we found that for large $r$ and small $n$,  $\om^{(r)}_{(n)} \propto b^{-r}$,  so that $\om_{(n+1)}/\om_{(n)} \to b$ asymptotically as $n\to 2$, with an accuracy of $\sim 10^{-4}$ to $10^{-3}$ for $r=9$ and $n=2,3,4$. In Fig. 1, Ref.\cite{Tuncer2015}, one cannot distinguish the $\om_{(n)}$ from equidistant points on a log-log plot. This is borne out by the asymptotic result we give in Eq.~(\ref{asymptotic}).

\section {The Eigenvectors}

In this section  we consider the eigenvectors spanning the eigenspaces associated with the series of eigenvalues $ \om^{(r)}_{(n)}\le 1$, with exponentially growing  degeneracies  
Eq.~(\ref{tau}). 
We will explicitly show how the eigenvectors  $\bu_{(2)}^{(r)}$ associated with the smallest nonzero eigenvalue for an $r$-tree are constructed. Then we will show that the eigenvectors $\bu_{(n)}^{(r)}$, $2 <n < r$, which live on subtrees rooted  at the $s=n-1$st generation of an $r$-tree 
are congruent to and can be  obtained from the eigenvectors $\bu^{r^{\prime}}_{(2)}$  living on  trees of size  
$2\le r^{\prime}\le r $. 
%
\begin{figure*}[!ht]
\includegraphics[width=5cm]{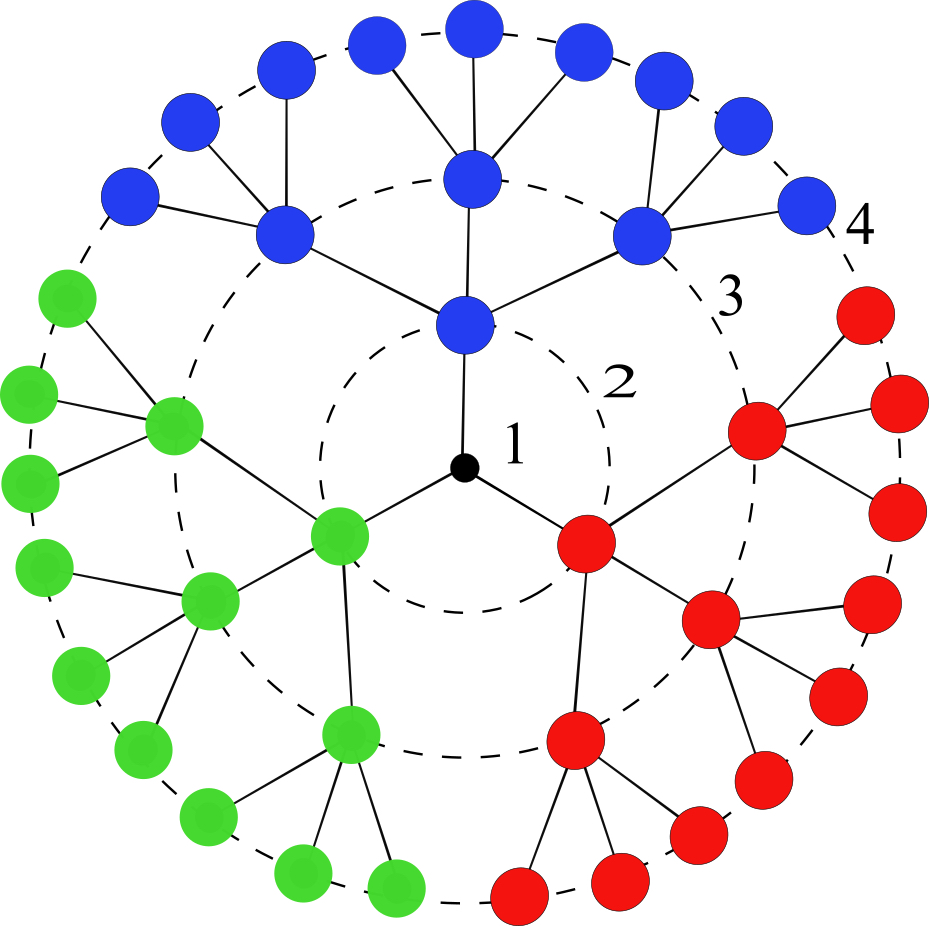}\;\;\;\;\;\;
\includegraphics[width=5cm]{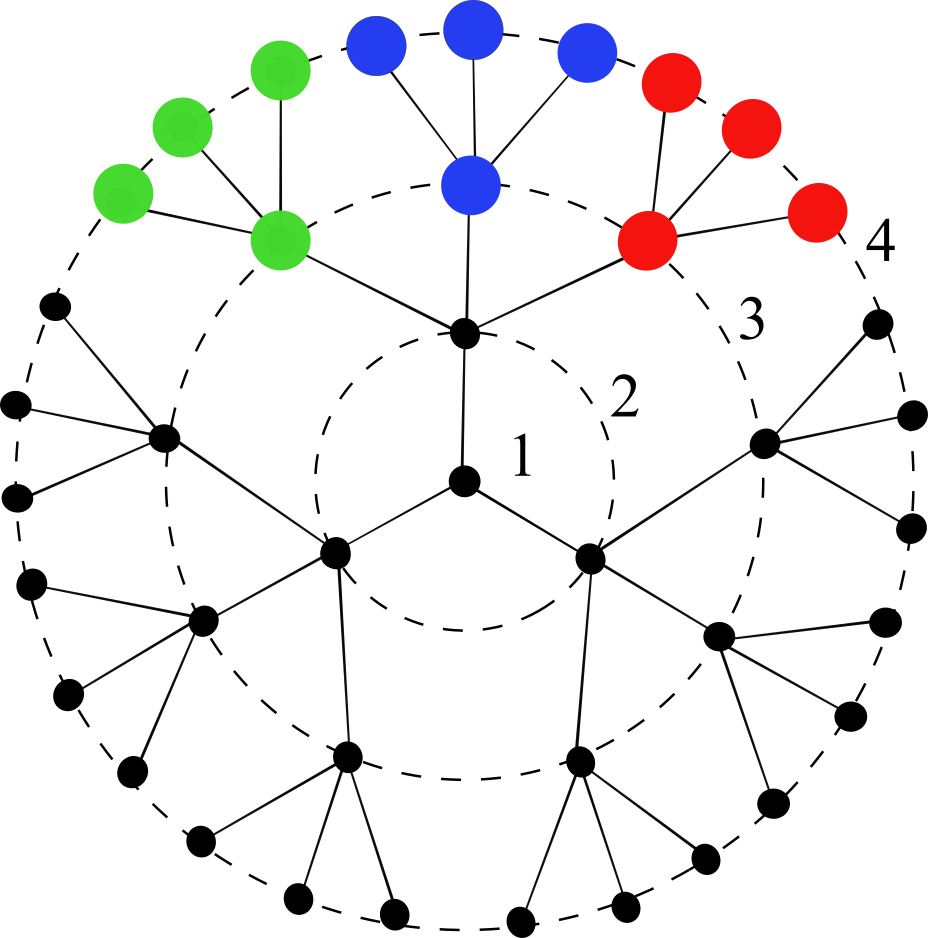}\;\;\;\;\;\;
\includegraphics[width=5cm]{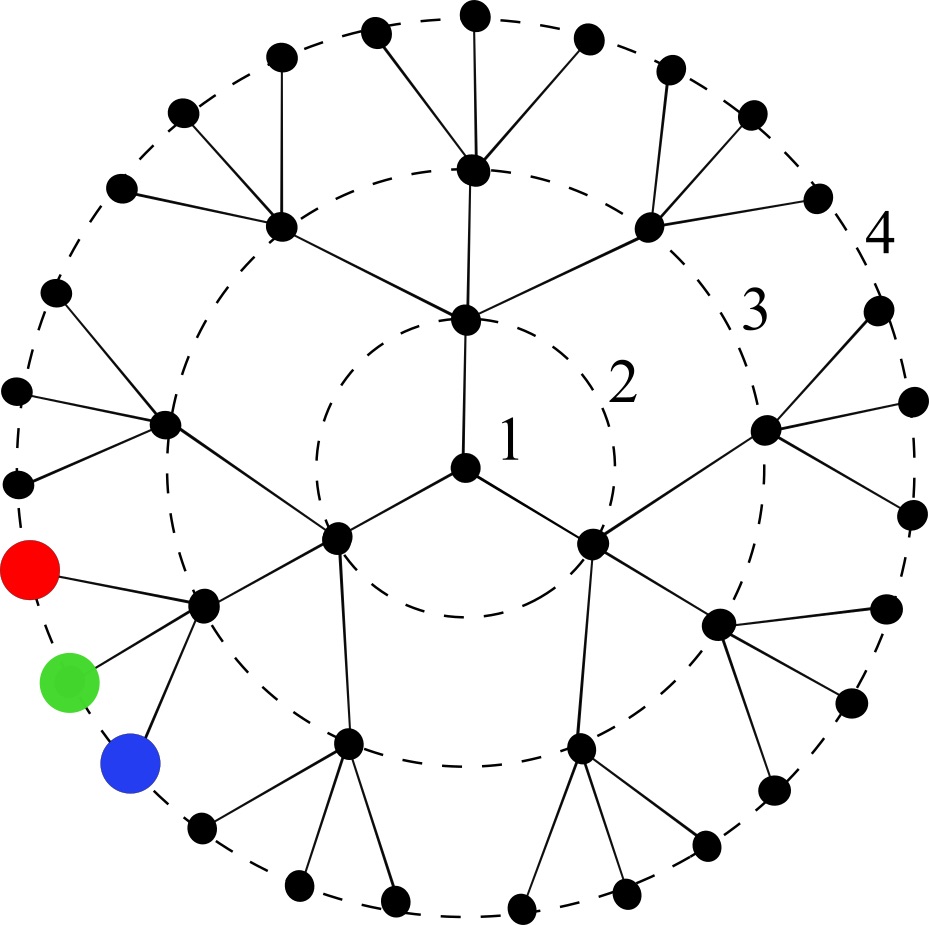}
\caption{(Color online) Cayley tree with four generations and branching number $b=3$.   We depict the eigenvectors $\bu_{1,(n)}^{(4)} ({\mathbf\rho})$,  belonging to the $n=2,3,4$th eigenspaces, in increasing order of their eigenvalues.  Black indicates  nodes corresponding to null elements of  the eigenvector, while blue, green and red correspond to the elements  $j =1,2,3$ of  $\bolde^{(\ell)}$.  In the first panel, the vector $\bu_{1,(2)}^{(4)}$ comprises the whole tree with four generations.  In the second and third panels, subtrees of size 3 and 2 are attached at 2nd and 3rd-generation nodes respectively. In the first two panels, $\ell=1$ while in the third, $\ell=2$, so that the green and blue have been interchanged. For a complete specification of the eigenvector we also need to specify the node at which the nontrivial subtree is attached. Since each node is connected to the root  (1st generation) by a unique path, the address of a node, say,  in the $s$th generation can be specified by a sequence of numbers $\mathbf{\rho}\equiv(1, \rho_2, \rho_3,\ldots  \rho_s)$, and  $1\le \rho_q\le b$, for all $q$. At the $q$th  step, the path crosses the $\rho_q$th node on the shell along its trajectory. In this figure, if we count clockwise starting from the left lower corner, the relevant subtrees are attached at the following nodes: ${\mathbf\rho} = (1);\;\;{\mathbf\rho}= (1,2); \;\; {\mathbf\rho}=(1, 1,2)$. (see Appendix 1).} 
 \label{trees}
\end{figure*}


\subsection{The eigenvector $\bu^{(r)}_{(2)}$ with the smallest nonzero eigenvalue}

It should be recalled that for any $r$, $\om^{(r)}_{(2)}$ is the smallest nonzero eigenvalue of the graph Laplacian and therefore its value is also called the ``spectral gap."

We make an ansatz that the eigenvector $\bu^{(2)} _{(2)}$ for the primitive tree provides a template for constructing the eigenvectors $\bu^{(r)} _{(2)}$ for the larger trees, with the first and second generations of the $r$-tree mimicking $\bu^{(2)} _{(2)}$ exactly.  The root node again has the value zero, as in the case of   $\bu_{(2)}^{(2)}$ (see Section II.2.); the values of the nodes belonging to the second generation (which consists of just one shell) are given by the elements of the vectors $\bolde^{(\ell)}$.   Now each subtree (of size $r-1$) emanating from a second generation node of the $r$-tree, uniformly inherits the complex value at that node, up to overall amplitudes $\zeta_m$ which differ from generation to generation. 

We now augment the vector $\bu_{(2)}^{(2)}$ such that it spans an $r$-tree.
Define $\boldb_m$ as,  
\be
\boldb_m\equiv \underbrace{\boldb\otimes \boldb\otimes \ldots \otimes \boldb}_\text{$m$ times}\;\;,
 \ee
with $\boldb_0 = 1$. The direct product $\bolde^{(\ell)} \otimes \boldb_{m-2}$ ``stretches"  $ \bolde^{(\ell)}$ across the whole extent  of the $m$th generation, with the first $b^{m-2}$ elements taking the value $e_1$, the second $b^{m-2}$ elements taking the value $e_2$, and so on up to the value $e_b$. This gives precisely $b^{m-1}$ nodes on the $m$th generation, as it should. 

The $m$th $b^{m-1}$-dimensional segment of the eigenvector, spanning the nodes of the $m$th generation on the $r$-tree, will be called a {\it subvector} and denoted by $\hat{u}_m$.  Each subvector is assigned an amplitude $\zeta_m$. For any given $r$, the proposed vector $\bu_{(2)}^{(r)}$ now takes the form,
\be \bu_{(2)}^{(r)} = \begin{pmatrix} 0\\ \zeta_2\, \bolde^{(\ell)} \otimes \boldb_{0} \, \\\ ..\\ \zeta_m\, \bolde^{(\ell)} \otimes \boldb_{m-2} \\ .. \\ \zeta_r\, \bolde^{(\ell)}\otimes \boldb_{r-2}\\ \end{pmatrix}
\equiv  \begin{pmatrix} 0\\ \hat{u}_2\\ ..\\ \hat{u}_m\\ .. \\ \hat{u}_r\\ \end{pmatrix}\label{ur}\;\;.
\ee

Now operating on this vector with the graph Laplacian, we obtain equations which we can solve for the  eigenvalue 
$\om_{(2)}^{(r)}$ and for the subvector amplitudes.

Requiring that ${\mathbf L}^{(r)} \bu^{(r)} = \om \, \bu^{(r)}$ and focusing on the $m$th subvector of the eigenvector $\bu^{(r)}_{(2)}$ we have, from Eq.~(\ref{Lfor1mr})for $1<m<r$,  
\begin{eqnarray}
\om^{(r)}_{(2)} \hat{u}_m &=  \sum_{n=1}^r {\mathbf L}^{(r)}_{m,n} \hat{u}_n \\ \nonumber
                          &= (b+1) \hat{u}_m -  {\boldB}_{m-2} \hat{u}_{m-1} -  {\boldB}^\dagger_{m-1} \hat{u}_{m+1} \;\;, \label{m}
\end{eqnarray}
a vector equality where the product ${\boldB}_{m-2} \hat{u}_{m-1}$ yields a vector with $b^{m-1}$ elements, and so does ${\boldB}^\dagger_{m-1} \hat{u}_{m+1}$. 
Note that  we have used  
\be
\boldB_1^\dagger \bolde^{(\ell)} \otimes \boldb  = b \,\bolde^{(\ell)}\;\;,
\label{blowup}
\ee
from Eq.~(\ref{Bnd}),  and
\begin{eqnarray}
{\boldB}_{m-2} \hat{u}_{m-1}&= [\boldI_b^{(m-2)}\otimes \boldb]\,\hat{u}_{m-1}  \\ \nonumber
                            & =  \hat{u}_{m-1} \otimes \boldb= \zeta_{m-1}\hat{u}_m / \zeta_m\;\;.
\label{note1}
\end{eqnarray}
Moreover, note that ${\boldB}_{m-2}$ is a $b^{m-1} \times b^{m-2}$ dimensional matrix which has the same effect on a vector $\mathbf{v}$ of dimension $b^{m-2}$ as  $ \mathbf{v}\otimes \boldb $, stretching it to a vector of dimension $b^{m-1}$ with each entry repeated $b$ times.

For the special case of $m=1$, $\hat{u}_1=0$ by assumption, and 
one has the trivial result, 
\be 
\om^{(r)}_{(2)}\hat{u}_1= 0 =0 - \zeta_2\, {\boldB}_0^\dagger\, {\mathbf e} = 0 \;\;
\ee 
since  ${\boldB^\dagger}_0 \bolde = \boldb^\dagger \bolde =\sum_{j=1}^b e_j=0$, with $e_j$ being the elements of 
$\bolde^{(\ell)}$; this is consistent with $ \hat{u}_1=0$
and we are therefore free to set $\zeta_1=0$. 

Let us do the   $m=2$ case explicitly. (We omit the superscript $(r)$ and $(\ell)$ from here till the end of this section to avoid clutter).
\be
\om_{(2)} \hat{u}_2= \om_{(2)} \zeta_2 \, \bolde  = 0 + \zeta_2 \,  \bolde (b+1)  -  \zeta_3 \, \boldB_1^\dagger \bolde \otimes  \boldb\;\;,
\ee
and using Eq.~(\ref{note1}), we get  
\be
\om_{(2)}\, \zeta_2 \,\bolde = \zeta_2(b+1) \,\bolde- \zeta_3\, b\, \bolde\,\,
\ee
or
\be
\zeta_2[\om_{(2)} -(b+1)] +b\, \zeta_3=0\;\;.
\label{new1}
\ee

In fact, for general $1 < m< r$,   from Eq.~(\ref{m}) 
and using the fact that 
\begin{eqnarray}
 \boldB_{m-1}^\dagger \hat{u}_{m+1} &=   \zeta_{m+1}[ \boldI_b^{(m-1)} \otimes \boldb^\dagger] \bolde\otimes \boldb_{m} \\
&=\zeta_{m+1}\, b\, \bolde \otimes \boldb_{m-1} \;\;,  \nonumber
\end{eqnarray}
with  $\bolde\otimes \boldb_{m-1}= \hat{u}_{m}/\zeta_m$, 
we get, after some simplification, 
\be
\om_{(2)}\hat{u}_m= -  \frac{\zeta_{m-1}}{\zeta_m} \hat{u}_m + (b+1)\hat{u}_m - \frac{\zeta_{m+1}}{\zeta_m} \hat{u}_m \;\;.
\ee
Dividing through by $\hat{u}_m$
we get the recursion relations for the amplitudes of the subvectors, 
\be
\om_{(2)}\zeta_m= -\zeta_{m-1}+(b+1)\zeta_m -b \zeta_{m+1} \;\;.
\label{recursionzeta}
\ee
Note that this equation holds for $m=2$ as well, since we have set $\zeta_1=0$.

For $m=r$,  
\be
\om_{(2)}  \hat{u}_{r}= 1 \zeta_r \bolde \otimes  \boldb_{r-2}\, -\, \zeta_{r-1} \boldB_{r-2}  \bolde \otimes  \boldb_{r-3}  \;\;, 
\ee
or,
\be
 \om_{(2)} \, \zeta_r \, = \zeta_r \, \bolde\otimes \boldb_{r-2}\,-\,  \zeta_{r-1}  \bolde \otimes  \boldb_{r-2}\;\;,
\label{sondanbirevvel}
\ee
which yields,
\be \om_{(2)} = (\zeta_r - \zeta_{r-1})/\zeta_r = 1-  \zeta_{r-1}/\zeta_r \;\;.
\label{son}
\ee
Choosing the relative amplitude $\zeta_r=1$ without loss of generality,  this yields $\om = 1-  \zeta_{r-1}$.  Thus we have exactly $r-1$ equations to solve for the $r-2$ unknown sub-amplitudes and  $\om_{(2)}$.  Substituting  $\om_{(2)}$  from Eq.~(\ref{son})  into Eq.~(\ref{sondanbirevvel}) and iterating the Eq.~(\ref{recursionzeta}) yields an $r-1$st order polynomial for the amplitudes $\zeta_m$, $m=2,\ldots,r-1$, as we demonstrate in the next subsection.  

\subsection{Solving the  recursion relations for the subvector amplitudes $\zeta_m$ }

The recursion relations for the subvector amplitudes turn out to be formally the same as those obtained for the eigenvalues of the graph Laplacian in the process of  solving the secular equation.
 We have already presented in Section II, how the solutions to 
difference equations like those in Eq.~(\ref{Arecursion}) may be found using the Cayley-Hamilton theorem~\cite{Kahn}. Here we will use this method and write a difference equation for the amplitudes  
$\zeta_m$ of the subvectors $\hat{u}_m$.

We may write Eq.~(\ref{recursionzeta}) as,
\be
\zeta_{m-1}=(b+1-\om_{(2)})\zeta_m-b\zeta_{m+1}\;\;,
\label{diffmatrix}
\ee
with the initial values $\zeta_r=1$ and $\zeta_{r-1}=1-\om_{(2)}$, and keeping in mind that we know $\zeta_1=0$.  This clearly has the form of Eq.~(\ref{Arecursion}).
The difference equation may be put in  matrix form,
\be
\begin{pmatrix} \zeta_{m-2} \\ \zeta_{m-1} \end{pmatrix} = \cal{A} \begin{pmatrix} \zeta_{m-1} \\ \zeta_{m} \end{pmatrix} \;\;.
\ee
where the matrix ${\cal A}$ is the same  as that defined in Eq.~(\ref{calA}).

Using the ansatz we have made for the vector $\bu^{(r)}_{(2)}$ and iterating down from the $r$th to the 2nd subvector amplitude and using the initial values given above, we have, 
\be 
\begin{pmatrix} 0 \\ \zeta_2 \end{pmatrix} ={\cal{A}}^{r-2}\begin{pmatrix}\zeta_{r-1} \\ 1 \end{pmatrix}  \;\;.
\label{zmatrix1}
\ee
The first equation we obtain from Eq.~(\ref{zmatrix1})) reduces, after some algebra, to
\be 
1- \om^{(r)}_{(2)} = b \left( \frac{\lambda_1^{r-2}- \lambda_2^{r-2}}{\lambda_1^{r-1}- \lambda_2^{r-1}}\right)\;\;.
\label{zsolution1}
\ee
This is exactly what we need to find  $\om^{(r)}_{(2)}$ since this eigenvalue is yielded by $A_{r-1} = 0$. Replacing $s$  in Eq.~(\ref{omsolution2}) by $(r-1)$ yields precisely the correct powers in the expression above.  

From Eq.~(\ref{zmatrix1}) we also obtain,
\be
\zeta_2=\alpha_{r-2} (1-\om^{(r)}_{(2)}) + \beta_{r-2} 
\label{zsolution2}
\ee
The rest of the amplitudes $\zeta_m$ can clearly be obtained by 
iterating Eq.~(\ref{diffmatrix}) upto $\zeta_r$. Note that for $m=1$, one has $b-\om_{(2)}$ rather than $\phi$ in Eq.~(\ref{diffmatrix})  since $\zeta_1=0$ by construction. 

%
\subsection{Eigenvectors $\bu_{(n)}^{(r)}$, $2<n < r$}
In the previous subsections we found that the eigenvector $\bu_{(2)}^{(r)}$ spans the whole tree, with a zero element located at the root (first generation) and nonzero elements with values given by elements of the vector $\mathbf{e}$ radiating from the second generation.
 
The eigenvectors spanning the eigenspaces of the successive nonzero eigenvalues  $\om_{(n)}^{(r)}$, $2 < n\le r$ 
belonging to the  ascending  series of eigenvalues are found in the following by   
shifting the root node of $\bu_{(2)}^{(r)}$ from the $s=1$ to the  to the $s=n-1$ generation and truncating the top $n-1$ generations which are now protruding beyond the $r$th generation. All the elements of the eigenvector, which are not determined in this way, are set  equal to zero.  

We illustrate this by an example, $\bu_{(3)}^{(5)}$. In this case the root of the vector $\bu_{(2)}^{(5)}$ must be moved to the second generation; we may pick any one of the $b$ nodes in this generation.  Let us specify $b=3$ and pick the second node. In this way the second generation of  $\bu_{(2)}^{(5)}$ will now reside within the 3rd generation of $\bu_{(3)}^{(5)}$, and since it emanates from the middle (i.e.,2nd) node in the second generation, it will occupy the middle shell in the third generation and so on. Defining the null vectors  $\mathbf{O}_n = 0 \otimes \boldb_n$, of size $b^n$, we have,
\be \bu_{\ell,(3)}^{(5)}(1,2) = \begin{pmatrix} 0\\ \zeta_1 \mathbf{O}_1 \\ \zeta_2 \begin{pmatrix} \mathbf{O}_1\\ \bolde^{(\ell)} \\ \mathbf{O}_1 \end{pmatrix} \\\ ..\\ \zeta_4\, \begin{pmatrix} \mathbf{O}_3\\  \bolde^{(\ell)}\otimes \boldb_{2}  \\\mathbf{O}_3 \end{pmatrix}  \end{pmatrix}
\ee
We use the  vector $\mathbf{\rho}=(1,2)$  to clearly specify where we root the subtree  $Y_{\mathbf{\rho}}^{r-1}$, on which the nonzero elements of the vector 
$\bu_{(n)}^{(r)}$ live. (see Appendix 1) However this information is already contained in the way we have inserted null vectors into our eigenvector $\bu_{\ell,(3)}^{(5)}$.  

Operating with the graph Laplacian for an $r=5$ tree on the vector $\bu_{\ell,(3)}^{(5)}$, we recover exactly the same equations and recursion relations for $\zeta_m$, as in Eqs.~(\ref{new1},\ref{recursionzeta},\ref{son},\ref{diffmatrix}), except that $1\le m \le 4$, since we have set the root equal to zero and assigned the subvector amplitudes $\zeta_m$ starting from the second generation.  Thus, the nonzero elements of $\bu_{\ell,(3)}^{(5)}$ are exactly those of  $\bu_{\ell,(2)}^{(4)}$, as we claim.

The degeneracies $\tau_{(n)}$ of the successive eigenvalues $\omega_{(n)}$ for $1<n\le r$ are immediately found, by noting that there are exactly $b^{n-2}$ sites on the $n-1$st generation, at which the subtrees with $r-n+2$ generations (counting the root of the subtree as the 1st  generation) can be attached. 
Within a given ($n$th) eigenspace, with the same wavenumber $(\ell)$, orthogonality is ensured since the subtrees with nonzero elements, attached at different sites,  do not overlap.  Clearly eigenvectors from the same eigenspace with $\ell \ne \ell^{\prime}$ are orthogonal to each other even if they occupy the same subtree.
Orthogonality of eigenvectors belonging to eigenspaces $ n^\prime \ne  n$, which may overlap, is ensured by the fact that their scalar product, which can be broken up into sums over shells, always involves scalar products of $\bolde$ with constant subvectors, the latter coming from an $n^\prime > n$th eigenspace. 

Let us take two eigenvectors from eigenspaces $2\le n <n^{\prime}$ (attached at the generations $s=n-1$ and $s^\prime=n^\prime-1$ respectively) without loss of generality.  For the non-zero elements of the eigenvectors to overlap, the first $n-1$ indices of the sites at which their respective subtrees are attached must be identical, i.e., $\rho_j = \rho^\prime_j$ for $j=1, \ldots, s$. In other words the subtrees $Y^{r-s^{\prime}}_{\rho_1,\ldots,\rho_{s^\prime}}$ must be a subtree of $Y^{r-s}_{\rho_1,\ldots,\rho_{s}}$.   Now ${\bu_{(n)}^{(r)}}^\dagger \bu_{(n^\prime)}^{(r)}$ involves sums over overlapping elements on the generations $s^\prime$ to $r$.  The $s^\prime$th generation is trivial since there is only one element of $\bu_{(n^\prime)}^{(r)}$ coinciding with the nontrivial subtree of $\bu_{(n)}^{(r)}$  on this generation, and it is zero. The elements of  $\bu_{(n)}^{(r)}$ on the generations $k> s^\prime$ are given by $\bolde \otimes \boldb_{k-s^\prime}$.  The range of the overlap is from the node   
\be
{\mathbf \rho} =(\rho_1,\ldots, \rho_{s^\prime-1},\underbrace{ 1,1,\ldots 1}_\text{$k-s^\prime$ times}
\ee
to the node
\be
{\mathbf \rho}= (\rho_1,\ldots, \rho_{s^\prime-1},\underbrace{ b,b,\ldots b}_\text{$k-s^\prime$ times}) \;\;.
\ee
Within this range  the elements of $\bu_{(n^\prime)}^{(r)}$  are $\bolde \otimes \boldb_{k-s^\prime}$, up to  the subvector amplitudes $\zeta_j$.

For the sake of illustration let us take $n^\prime = n+1$, and the subtrees $Y^{r-s+1}_{\rho_1,\ldots,\rho_{n}}$ and $Y^{r-s}_{\rho_1,\ldots,\rho_{n},k}$.  Then,on the $s$th generation,
\be
\big [{\bu_{(n^\prime)}^{(r)}}^\dagger \; \bu_{(n^\prime)}^{(r)}\big] \propto e_{k} \cdot 0=0\;\;,
\ee
and on the $n+1$st generation,
\be 
\big[{\bu_{(n)}^{(r)}}^\dagger \bu_{(n^\prime)}^{(r)}\big] \propto e_{k}  \boldb^\dagger  \bolde = e_{k}  \sum_{j=1}^b e_j = 0 \;\;.
\ee
On all the succeeding generations $s$, with  $s-s^\prime=m$, we will have contributions to the scalar product in the form 
\be  \boldb_m^\dagger  \big[\boldb_{m-1} \otimes \bolde \big]\propto  \sum_{j=1}^b e_j  b^{m-1} = 0 \;\;.
\ee
In fact, this will be the case  for any $n^\prime-n\ge 1$, since the overlap is determined by $n^\prime$ only.

The set of eigenvectors which we have presented here may of course be combined in different ways in contexts for  which it may be more convenient to stress other aspects of the eigenspaces.

\section{Discussion}

This research was initiated in order to widen the range of our tools for analysis on arbitrary networks.  In particular, we wished to understand the structure of the eigenvectors of the graph Laplacian on such a simple network as the Cayley tree so that we would eventually be able to use these eigenvectors in generalised Fourier transformations of fields living on the nodes of the network.  In a previous paper~\cite{Tuncer2015}, where we introduced a ``field theoretic," Wilson-style renormalization group scheme, we made use of numerically computed eigenvectors for a Cayley tree. However, the numerics obscured the symmetry properties of the network. We believe that obtaining the eigenvectors analytically, by making use of some insight and the symmetries of this much exploited network, has turned out to be a useful pedagogical exercise. 

It is interesting to see that in the small eigenvalue ($0\le \om \le 1$) region of the Laplace spectrum, one finds eigenvectors which pick out large scale features spanning many generations, as well as those which zoom in  on  just a shell  at the tip of the tree, attached to a node on the $r-1$st generation. For a tree with uniform branching number $b$, the articulation in the transverse direction (i.e., staying within any one generation) involves complex exponentials $e^{i 2\pi j\ell/b}$, where  $j=1,\ldots,b$ and $\ell = 1, \ldots, b-1$. The eigenvalues can be found as the zeroes of a set of polynomials $A_s$, $s=1, \ldots\ r$, which are nested within each other, just as the nonzero elements of the eigenvectors are nested within each other, occupying smaller and smaller trees for larger and larger eigenvalues in the interval $0\le \om\le 1$. 

The radial symmetry of the Cayley tree suggests a similarity with a discrete Bessel equation and the Bessel functions, which deserves further attention.

{\bf Acknowledgements}

Ay\c se Erzan is a member of the Bilim Akademisi (Science Academy, Turkey).
Asl\i  Tuncer acknowledges partial support from the Istanbul Technical University Scientific Research Projects fund  ITU BAP  36259. 

\appendix
\renewcommand{\thefigure}{A\arabic{figure}}
\setcounter{figure}{0}
\vspace*{1cm}
\begin{figure*}[!ht]
\begin{center}
\includegraphics[width=16cm,trim={1cm 0 0 0}]{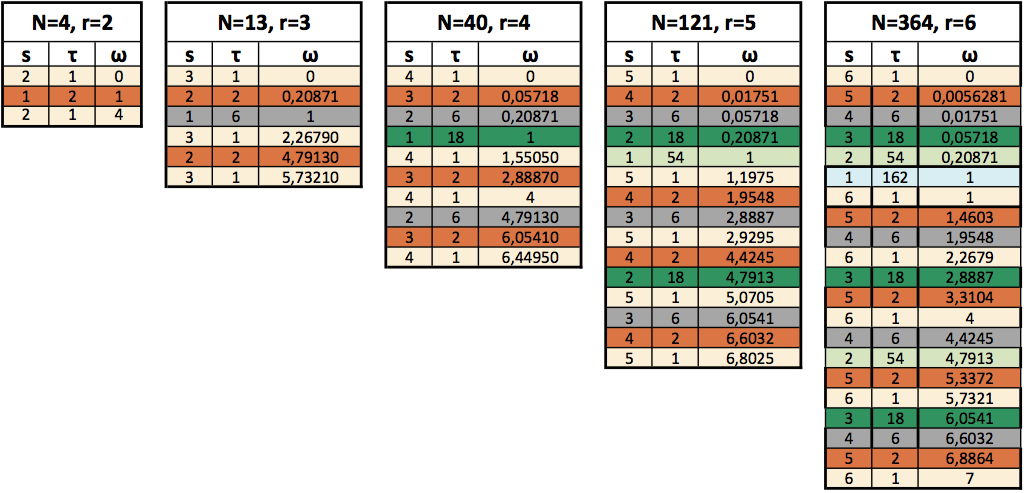}
\caption{(Color online) All eigenvalues of trees with $r=1, \ldots\ 6$, with branching number $b=3$. Clearly, setting the polynomials of degree $s$, 
Eq.~(\ref{D2}), $A_s=0$ yields $s$ distinct eigenvalues. $A_r$ has degeneracies $\tau=1$. For $1\le s \le r-1 $,  we simply get 
$\tau=(b-1)b^{r-s-1}$.  Eigenvalues with identical degeneracies have been highlighted with the same color.} 
\end{center}
\label{trees}
\end{figure*}

{\bf Appendix 1: Generating  and labelling the nodes of the Cayley tree}
Let us symbolically denote a primitive tree with $b$ branches by $Y(b)$.  We define a multiplication rule $(\ast)$ for trees such that if $Y$ and $X$ are trees, $Y\ast X$ is a tree which has $Y$  attached by its root to each ``leaf" of $X$.  Then, $Y(b)^{r-1}=Y(b)\ast Y(b)\ast \ldots \ast Y(b)$ with $r-1$ factors, is the finite uniform Cayley tree of $r$ generations (or of size $r$). 

A standard way to label the nodes of the $r$-tree is to assign $n$ indices to each node on the $n$th generation of the tree, {\it e.g.}, $\{\rho_1, \rho _2, \ldots, \rho _n\}$, where each $\rho _k$ can take values $1,2,\ldots, b$. In this way the site can be located by making the appropriate choice among each of the $b$ branches as one travels from the root to a site on the $n$th shell.   The root has only one possible address, $\rho_1=1$ and is therefore often omitted from the genealogy of the other nodes. However, we have kept this index as well. Thus the first node on the first generation (the root) thus has $\rho _1=1$ and the last node on the last generation on an $r$-tree has the address $\rho_1=b, \rho_2=b, \ldots, \rho_r=b$. 

Let us now consider subtrees origination from a site on the $p$th generation of an $r$-tree.  Such a subtree will have a total number of  $r-p+1$ generations, including the site from which it originates. The number of different sites (from which it can originate) on the $p$th generation is $b^{p-1}$. We need to specify precisely $p-1$ numbers to determine the site on the $p$th generation where this subtree is to be  attached, each number indicating a node (out of  $b$ nodes) lying on the path which eventually ends on  the site in question. This subtree will be then be labelled $Y^{r-p}_{\rho_1, \ldots, \rho_{p-1}}$.

{\bf Appendix 2: The eigenvalues with their attendant polynomials}

In Fig.~\ref{trees} below we provide tables of all the eigenvalues for trees of size 2 to 6. The eigenvalues (including the null eigenvalue) found by setting $A_r=0$ are non-degenerate.  For $1\le n \le r$, the leading $n$th eigenvalue corresponds to the smallest solution of   $A_{r-n+1} = 0$. Within this interval, the degeneracies $\tau_{(n)}$ are given by $b^{n-2}(b-1)$, where the first factor is equal to the number of nodes on the  $n-1$st generation, and the second factor from the ``wave number" $\ell = 1, \ldots, b-1$. (Also see Eq.~(\ref{D2}).)  The degeneracies $\tau$ for the whole spectrum can in fact be read off  from Eq.~(\ref{D2}), since, regardless of the order ($n$) in which the eigenvalues appear, the polynomials $A_s$, with $1 \le s\le  r-1$  occur with the powers $(b-1) b^{r-(s+1)}$, which yield the degeneracies of the roots.

\end{document}